\documentclass[aps,prl,twocolumn,superscriptaddress]{revtex4-1}
\usepackage{graphicx}
\usepackage{units}
\begin{document}
\title{Tuning effective hyperfine fields in PEDOT:PSS thin films by doping}
\author{M. Y. Teferi}
\affiliation{Department of Physics and Astronomy, University of Utah, Salt Lake City, Utah 84112, USA}
\author{J. Ogle}
\affiliation{Department of Chemistry, University of Utah, Salt Lake City, Utah 84112, USA}
\author{G. Joshi}
\author{H. Malissa}
\author{S. Jamali}
\author{D. L. Baird}
\affiliation{Department of Physics and Astronomy, University of Utah, Salt Lake City, Utah 84112, USA}
\author{J. M. Lupton}
\affiliation{Institut f\"{u}r Experimetalle und Angewandte Physik, Universit\"{a}t Regensburg, Resensburg, Germany}
\author{L. Whittaker Brooks}
\affiliation{Department of Chemistry, University of Utah, Salt Lake City, Utah 84112, USA}
\author{C. Boehme}
\affiliation{Department of Physics and Astronomy, University of Utah, Salt Lake City, Utah 84112, USA}
\date{\today}
\begin{abstract}
Using electrically detected magnetic resonance spectroscopy, we demonstrate that doping the conducting polymer poly(3,4-ethylenedioxythiophene):poly(styrene-sulfonate) (PEDOT:PSS) with ethylene glycol allows for the control of effective local charge carrier hyperfine fields through motional narrowing. These results suggest that doping of organic semiconductors could enable the tuning of macroscopic material properties dependent on hyperfine fields such as magnetoresistance, the magneto-optical responses and spin-diffusion.
\end{abstract}
\maketitle
The omnipresence of hydrogen in organic semiconductor thin film materials causes highly localized and randomly oriented magnetic hyperfine fields of varying effective strength \cite{PhysRevLett.104.017601,malissa2014room}. These hyperfine fields can crucially affect a variety of electronic, opto-electronic and magneto-electronic properties of organic semiconductor materials, including magnetoresistance \cite{nguyen2010isotope,waters2015spin}, spin-diffusion lengths \cite{sun2016inverse}, spin-relaxation times \cite{Phys.Rev.Lett.108.267601.2012}, and the efficiency of electroluminescence in organic light emitting diodes (OLEDs) \cite{lee2011tuning,kavand2016discrimination}. Thus, the ability to engineer hyperfine fields for specific technological applications has become important. Most straight forwardly, hyperfine field engineering is possible through hydrogen isotopic substitution \cite{nguyen2010isotope,lee2011tuning,malissa2014room}. However, as deuterium is the only stable hydrogen isotope next to protium, the range of achievable hyperfine field distributions using this approach is limited. In addition, deuteration in organic synthesis can be very difficult to achieve depending on the availability of starting materials.

In this letter we pursue the demonstration of a different approach to control charge carrier hyperfine fields, namely through tuning of motional narrowing. Motional narrowing reduces effective charge carrier hyperfine field strengths through time-domain averaging of local hyperfine fields across the various positions that a charge carrier passes as it propagates through a given material \cite{Phys.Rev.Lett.108.267601.2012}. Fig.~1 illustrates this effect by depicting the two cases of slow and fast charge carrier hopping. In the slow hopping regime, the average value for hopping transition (or dwell) times  significantly exceeds the precession period of the charge carrier spin as determined by the broad distribution of local hyperfine fields. Consequently, fast randomization of the charge carrier's spin states occurs during transport. In contrast, when hopping is fast, much reduced spin precession will take place at the individual hopping sites and therefore the spins randomize very slowly. This effect is commensurate with a very small, motionally narrowed, effective distribution of hyperfine fields \cite{Phys.Rev.Lett.108.267601.2012}.

The existence of motional narrowing in organic semiconductors was suggested in the past \cite{PhysRevLett.105.176601} for the $\pi$-conjugated polymer (poly[2-methoxy-5-(2-ethylhexyloxy)-1,4-phenylenevinylene], (MEH-PPV). However this conclusion did not withstand further scrutiny \cite{boehme2013challenges,kupijai2015bipolar}, likely due to insufficiently slow charge carrier hopping transitions \cite{Phys.Rev.Lett.108.267601.2012}. However, motional narrowing has been observed in high-mobility organic semiconductors in particular in field effect transistors of pentacene \cite{PhysRevLett.100.126601,Jpn.J.Appl.Phys.48.04C175.2009,PhysRevB.87.045309} and thienothiophene-based high-mobility materials \cite{PhysRevB.84.081306,PhysRevB.85.035308,APL105.033301} through temperature-dependent narrowing of the continuous wave electron paramagnetic resonance (EPR) line of field-induced charge carriers at X-band frequencies. We therefore have focused in this study on the conducting polymer blend poly(3,4-ethylenedioxythiophene):poly(styrene-sulfonate) (PEDOT:PSS). PEDOT:PSS is a technologically important polymeric system which has excellent thermal stability, solution processability, mechanical flexibility, and, most significantly for this study, relatively high charge carrier mobilities as compared to other organic conductors and semiconductors and, thus, high electrical conductivity \cite{PrincipPolymer2010}. PEDOT:PSS is widely used as a thin film electrode in organic-based electronic devices, as a transparent thin-film conductor for antistatic coatings, and in display applications \cite{PrincipPolymer2010}. Many research efforts have been focused on increasing the electrical properties of PEDOT:PSS via doping \cite{Adv.Electron.Mater.1.1500017.2015} with molecules like ethylene glycol (EG), dimethyl sulfoxide, sorbitol, N,N-dimethyl formamide, and diethylene glycol. All these materials used as doping agents have been shown to increase the electrical conductivity of PEOT:PSS by more than three orders of magnitude \cite{Adv.Electron.Mater.1.1500017.2015}, an effect predominantly attributed to mobility increase. This increase has been suggested to be due to either solvent-induced PEDOT:PSS improvements of interchain ordering, i.e. microscopic ordering, or due to conformational changes \cite{Adv.Mater.25.2831.2013, Adv.Funct.Mater.15.203.2005}.

In the following, we show that the high mobilities of PETDOT:PSS at low temperature cause montional narrowing of charge carrier hyperfine fields and that the ability to tune mobilities using EG \cite{Adv.Mater.25.2831.2013} also enables the tunability of effective charge carrier hyperfine fields and associated observables such as charge carrier spin coherence times. In order to detect motional narrowing, we measured the distribution of local effective hyperfine field strengths as well as the spin decoherence times ($T_2$) using continuous wave (cw) \cite{lee2012modulation} and pulsed (p) \cite{boehme2017electrically} electrically detected magnetic resonance (EDMR) spectroscopies. In order to detect EDMR, changes of spin-dependent currents of paramagnetic charge carriers are measured while the charge carrier spin states are brought into resonance \cite{lee2012modulation}.

Following the analytical approach outlined by Baker et al. \cite{Phys.Rev.Lett.108.267601.2012}, the correlation of local effective hyperfine field strengths and the spin decoherence times $T_2$ unambiguously reveals the presence or absence of motional narrowing: while an increase of $T_2$ with increasing mobility indicates motional narrowing of the effective hyperfine field distributions, $T_2$ will decrease with increasing mobility in the slow hopping regime where motional narrowing is absent. Thus, an increase in $T_2$, together with the narrowing of the EDMR line, is characteristic for motional narrowing.

\begin{figure}
\label{Fig1}
\includegraphics[width=\columnwidth]{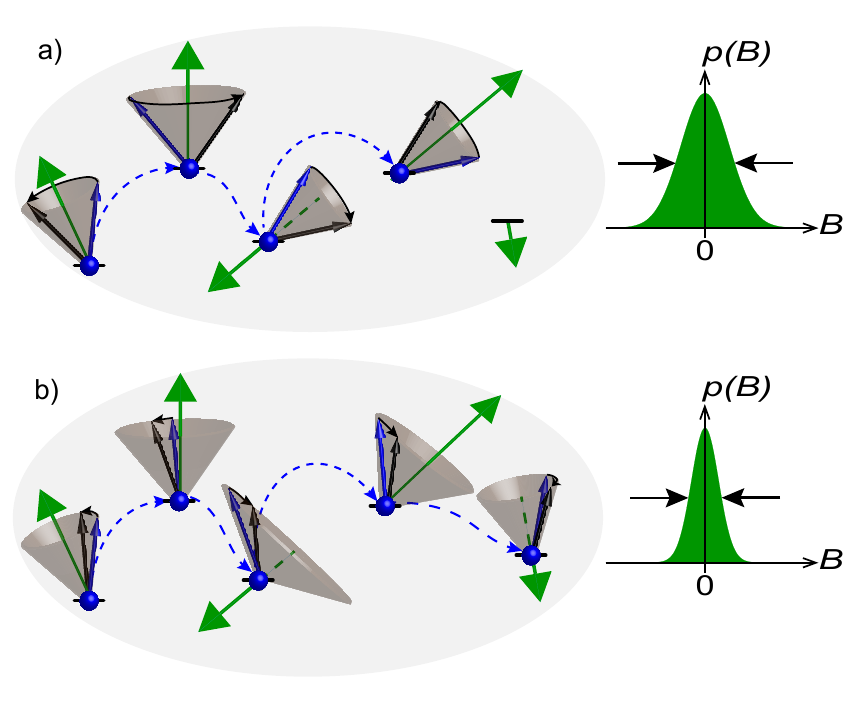}
\caption{Illustrations of spin randomization of charge carriers's spin state while the carrier is undergoing hopping transitions through localized paramagnetic charge carrier states. The green arrows represent local random hyperfine field directions at the different hopping sites while the blue and black arrows represent the carrier's spin states for the times when the charge transitions into and out of the localized states, respectively. The dashed curved arrow indicates the hopping. (a) shows the case of slow hoping with the effective distribution of nuclear magnetic fields $p(B)$ sketched on the right. When fast hopping occurs (b), the charge carrier spin state precesses only marginally before the next hopping transition occurs, implying that the distributions of the effective hyperfine fields acting on the charge carrier spins are considerably narrower, i.e. motionally narrowed.}
\end{figure}

 Experimentally, we prepared a series of $\approx$\unit[50]{nm} thin films by spin coating aqueous solutions containing PEDOT:PSS (Clevios P VP AI 4083) that were doped with EG with various weight concentrations between 0\% and 0.125\%. The films were deposited on EDMR compatible device templates which consisted of lithographically patterned indium tin oxide (ITO) bottom electrodes on glass substrates as described elsewhere \cite{J.Appl.Phys.114.133708}. After the EG doped PEDOT:PSS film deposition, the devices were capped with thermally evaporated aluminum thin films as top electrodes. The active area for all devices was set to \unit[2]{mm} $\times$ \unit[3]{mm}.

\begin{figure}
\includegraphics[width=\columnwidth]{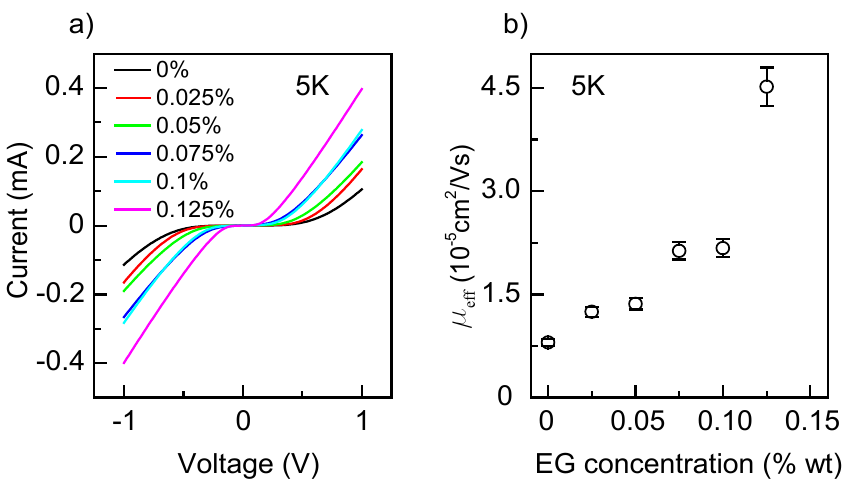}
\caption{Electrical characterization of the PEDOT:PSS thin-film devices for different EG doping concentrations. (a) Plot of the IV characteristics at \unit[5]{K} for various
EG concentrations. The observed functions are non-Ohmic, yet overall, the current still increases with doping concentration. (b) Plot of the low temperature charge carrier mobilities that were obtained by fits of Mott-Gurney's law to the devices' low-temperature IV characteristics, as a function of the applied EG concentration.}
\end{figure}

For each device, the current-voltage (IV) characteristics were recorded at room temperature (RT), verifying the expected Ohmic behaviors with decreasing resistance. The IV characteristics was then also recorded at $T=\unit[5]{K}$, yielding non-Ohmic behaviors as shown in Fig.~2(a), that were governed by space-charge limited transport. This allowed for the determination of majority charge carrier mobilities in a procedure based on the fit of Mott-Gurney's law as discribed in the supplementary information \cite{SI}. The obtained mobilities are displayed in Fig.~2. After the electrical characterization, multifrequency cwEDMR and Hahn-echo decay detected pEDMR experiments were carried out at $T=\unit[5]{K}$ in order to determine the effective magnetic hyperfine field distributions of the charge carriers as well as their transverse spin-relaxation times.

\begin{figure}
\includegraphics[width=\columnwidth]{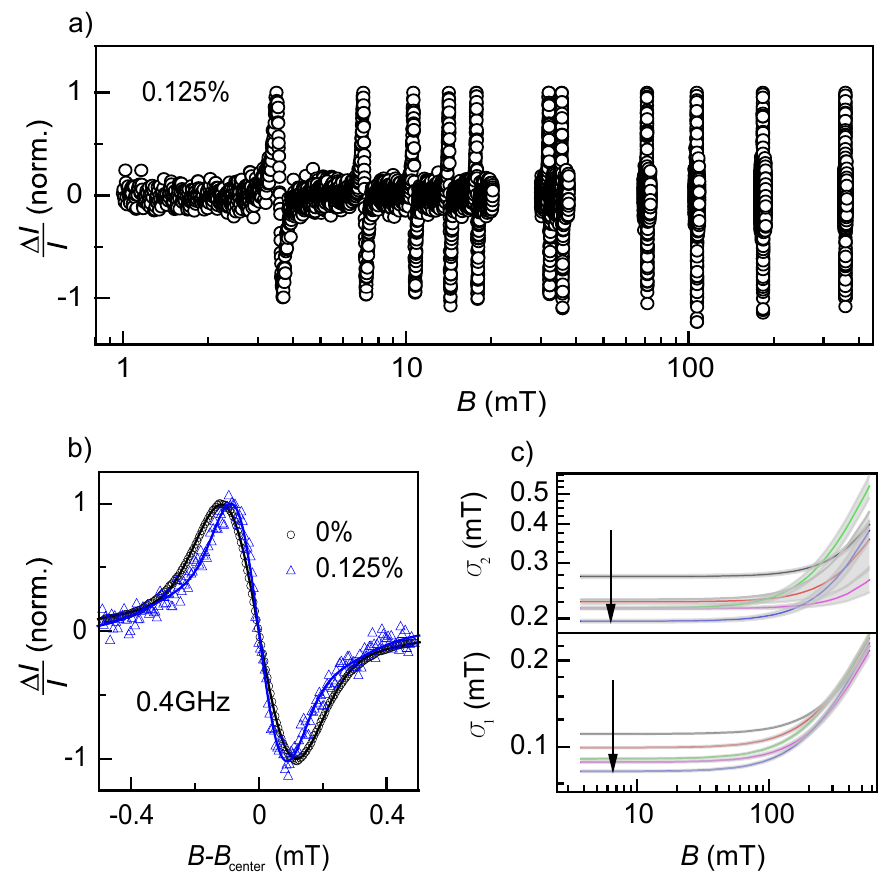}
\caption{(a) Results of magnetic field modulated, lock-in detected, cw-EDMR measurements of PEDOT:PSS films with 0.125\% EG doping at $T=\unit[5]{K}$ under application of various excitation frequencies between \unit[100]{MHz} and \unit[17]{GHz}. (b) Plot of two normalized spectra measured at $f=\unit[0.4]{GHz}$ for  0\% (undoped) and 0.125\% EG doped PEDOT:PSS samples. A derivative double-Gaussian function is fitted to the data to estabilish the line width for each carrier species in the pair. (c) Plots of the widths $\sigma_{1}$ and $\sigma_{2}$ of the two charge carrier states in PEDOT:PSS as a function of the applied magnetic fields. These plots are based on fit results of the measurement series displayed in (a), acquired for all doping concentrations studied. The arrows indicate increasing EG concentration. The gray areas indicate uncertainy domains for the respective curves.}
\end{figure}
The detection of motional narrowing and an unambiguous measurement of local hyperfine field distributions, which are determined by magnetic resonance lines of paramagnetic charge carrier states, is possible as long as influences by other inhomogeneous broadening effects such as spin-orbit coupling (SOC) induced Land\'e $g$-factor distributions (the so-called $g$-strain) can be neglected. Hyperfine broadening of EPR lines is inherently independent of the magnitude of the magnetic field $B$ under which EPR spectra are recorded, while the width of $g$-strain and $g$-anisotropy \cite{PhysRevB.84.081306,PhysRevB.85.035308,APL105.033301} broadening is proportional to $B$ \cite{Appl.Phys.Lett.109.103303.2016,Rev.Sci.Instrum.87.1131106.2016}. The spectral line width determined at one single frequency does therefore not reveal both broadenings directly. Thus, in order to identify magnetic field regimes where the spectral broadening is independent of magnetic field, we subjected all samples studied to a series of cwEDMR measurements conducted at many frequencies. This is achieved by using coplanar waveguide (CPW) resonators, following a procedure described by Joshi et al. \cite{Appl.Phys.Lett.109.103303.2016}.

These measurements took place under a constant bias that was adjusted for each device to yield equal device currents of $\unit[50]{\mu A}$. The external magnetic field was  modulated at a frequency of \unit[500]{Hz} and an amplitude $B_m$ that was verified to be small enough such that modulation broadening of the resonance peaks could be excluded following the procedure described in Ref.~\onlinecite{J.Appl.Phys.35.1217.1964}.

Figure 3(a) shows a plot of the normalized current changes of cwEDMR spectra at various excitation frequencies for a 0.125\% EG doped sample. This multifrequency measurement series was repeated for all EG concentrations studied. The measurements showed that the EDMR signal magnitude decreases with increasing EG concentration as expected when an increased mobility due to shorter transition times between localized charge carrier states causes the lifetime of the intermediate spin-pair responsible for the EDMR signal in PEDOT:PSS \cite{Nature.Commun.6.6688.2015} to decrease. In fact, the restriction of this study to EG doping concentrations below 0.125\% was made as EDMR signals disappear beyond this concentration, likely because of the fast hopping times resulting from the increased mobility. Fig.~3(b) compares the cwEDMR spectra of two doping levels at at resonance frequency of $f=\unit[0.4]{GHz}$. Following the procedure described by Joshi et al. \cite{Appl.Phys.Lett.109.103303.2016}, each set of multifrequency cwEDMR spectra was subjected to a global fit which uses only six parameters in order to simultaneously fit all resonance lines and lineshapes. These parameters are the two widths of the Gaussian hyperfine field distributions of each carrier in the pair, i.e. presumably electron and hole \cite{Nature.Commun.6.6688.2015}, the two widths of the associated SOC induced Land\'e $g$-factor distributions, as well as the centers of the two magnetic  resonance lines. Furthermore, for each global fit, a statistical boostrap analysis, as outlined in detail in Ref. \onlinecite{Appl.Phys.Lett.109.103303.2016}, was performed in order to obtain the statistical variation of each fit parameter. The entire fits results can be used to plot predictions of the resonance line broadening for both carriers of the pair, as a function of magnetic field. These values are plotted in Fig.~3(c) (black line), together with the 95\% uncertainty intervals, as obtained from a so-called bootstrap procedure \cite{Appl.Phys.Lett.109.103303.2016,PhysRevB.94.214202,PhysRevB.97.161201}, represented by the gray areas. This field dependence shows that the contributions of SOC-induced line broading of the constituents line widths $\sigma_1$ and $\sigma_2$ dominate at high magnetic fields while the field-independent hyperfine broadening is dominant at low magnetic fields. The plateau values of $\sigma_1$ and $\sigma_2$ found in Fig.~3(c) give a direct and accurate (due to the very small uncertainty intervals) measure of the hyperfine field distributions. These values are plotted in Fig.~4(a) as a function of the EG doping concentration. The plots show that the line width decreases with increasing EG concentration and thus with increasing effective charge carrier mobility, as expected for motional narrowing.
\begin{figure}
\includegraphics[width=\columnwidth]{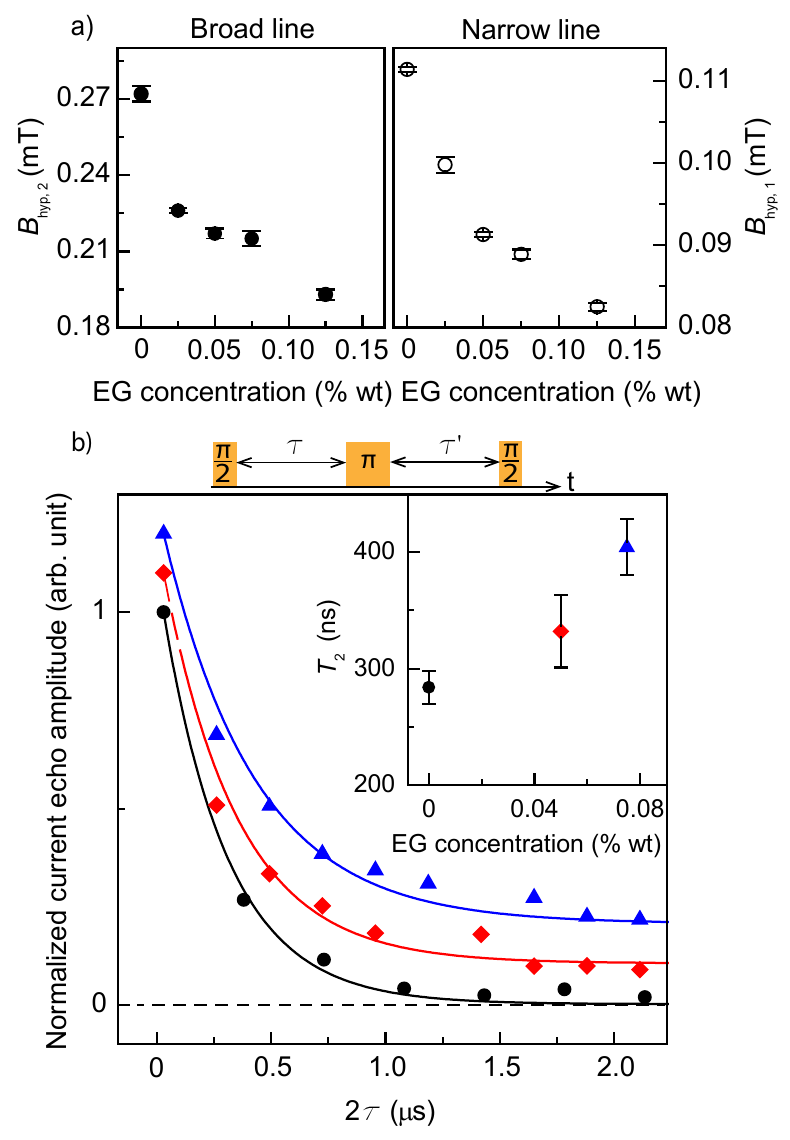}
\caption{ a) Plot of hyperfine field distribution width versus EG concentration for the broad and narrow Gaussian lines of EDMR spectrum . b) Plot of normalized charge carrier spin echoes versus interpulse delay for three different EG concentrations. The curves are vertically offset for clarity. Solid lines show the fit. Inset: the coherence time $T_2$ versus EG concentration.}
\end{figure}

The narrowing effect of charge carrier hyperfine field distributions with increasing mobility is entirely consistent with the motional narrowing hypothesis. Nevertheless, this observation alone is not an unambiguous proof since conformational changes within the PEDOT:PSS films could also cause changes of local hyperfine fields due to a proton rearrangement. We therefore further scrutinized the hypothesis by measuring the charge carrier spin-coherence times for the different doping concentrations. In order to do this, we carried out a series of electrically detected Hahn-echo experiments.

Spin coherence times were measured by applying a series of a modified three-pulse $\frac{\pi}{2}$-$\tau$-$\pi$-$\tau'$-$\frac{\pi}{2}$ Hahn-echo sequence consisting of the conventional two-pulse Hahn-echo sequence with $\pi$-pulse lengths of \unit[32]{ns}, followed by a delay time $\tau'$ prior to a subsequent $\frac{\pi}{2}$ projection pulse. The delay time was varied such that $|\tau'-\tau| < 16$ ns, following the EDMR detected Hahn-echo measurement protocol explained elsewhere \cite{PhysRevLett.100.177602,PhysRevB.81.075214}. Fig.~4(b) shows the results of electrically detected charge carrier spin echoes, measured with pEDMR, as a function of twice the interpulse separation ($2\tau$). The intensities of the observed echoes decrease with increasing $\tau$, indicating the gradual decay of the charge carrier spin ensembles. The spin decoherence times $T_2$, also known as transverse spin-spin relaxation times, are extracted from exponential fits, while the confidence intervals are obtained by statistical bootstrap analysis \cite{Ann.Stat.7.1.1979,PhysRevB.97.161201}. The results of these procedures, plotted in the inset of Fig.~4(b) as a function of increasing EG concentration, show that $T_2$ increases with EG concentration. Thus, taking the data of Figs.~2 and 3 into account, $T_2$ increases with increasing mobility and decreasing hyperfine broadening. On the one hand, this observation is expected, since carrier hopping in the inhomogeneous distribution of hyperfine fields constitutes the dominant decoherence mechanism \cite{Phys.Rev.Lett.108.267601.2012}. However, it is precisely this spin mixing by spin precession in the hyperfine fields which is required to generate a pEDMR signal and which probes spin-dependent transport and recombination. As a consequence, as the mobility increases, raising $T_2$, the observed magnitude of the EDMR signal decreases---a non-intuitive observation.

In conclusion, the observations presented above have demonstrated that (i) motional narrowing, as determined from the multi-frequency cwEDMR line width, as well as the spin decoherence time $T_2$, determines effective charge carrier hyperfine fields of PEDOT:PSS and (ii) these hyperfine fields are tunable by EG dopants which changes the effective mobility. We note that the effective narrowing observed in this study ($\approx$30\% and $\approx$23\% for the two charge carrier species) are solely limited by the applicability of the EDMR method. Narrowing likely continues when EG doping exceeds 0.125\% and will likely also occur for other PEDOT:PSS dopants and in other high-mobility organic conductors and semiconductors. Thus, doping-induced tuning of effective charge carrier hyperfine fields can provide a pathway for tuning spin-coherence times and spin-diffusion lengths, magnetoresistance and other hyperfine-field dependent magneto-optoelectronic materials characteristics.
\begin{acknowledgments}
We acknowledge support of M. Y. T and J. O. through the NSF Material Research Science and Engineering Center at the University of Utah (DMR-1121252). G. J., H. M., and S. J., were supported by the Department of Energy (DE-SC0000909). The authors thank Richards Miller III for helpful discussions on the statistical bootstrap analysis.
\end{acknowledgments}
\bibliography{References_cmb_comp}
\end{document}